\documentclass{svproc}
\usepackage{tikz}
\usetikzlibrary{automata, positioning, arrows, arrows.meta}
\usepackage{pgfplots}
\pgfplotsset{compat=1.17}
\usepackage{graphicx}%
\usepackage{multirow}%
\usepackage{amsmath,amssymb,amsfonts}%
\usepackage{mathrsfs}%
\usepackage[title]{appendix}%
\usepackage{xcolor}%
\usepackage{textcomp}%
\usepackage{manyfoot}%
\usepackage{booktabs}%
\usepackage{algorithm}%
\usepackage{algorithmicx}%
\usepackage{algpseudocode}%
\usepackage{listings}%
\usepackage{caption}
\usepackage{subcaption}
\usepackage{url}
\usepackage{enumitem} 

\begin{document}
\mainmatter              
\title{DynamicScore: a Novel Metric for Quantifying Graph Dynamics}
\titlerunning{DynamicScore}  
%
\author{Bridonneau Vincent\inst{1} \and Guinand Frédéric\inst{1} \and
Pigné Yoann\inst{1}}
\authorrunning{Bridonneau Vincent et al.} 

\institute{LITIS Lab, Université Le Havre Normandie \\
\email{firstname.lastname@univ-lehavre.fr}
}

\maketitle              

\begin{abstract}
This study introduces a new metric called “DynamicScore” to evaluate the dynamics of graphs. 
It can be applied to both vertices and edges. 
Unlike traditional metrics, DynamicScore not only measures changes in the number of vertices or edges between consecutive time steps, but also takes into account the composition of these sets. 
To illustrate the possible contributions of this metric, we calculate it for increasing networks of preferential attachment (Barabási-Albert model) and Edge-Markovian graphs. 
The results improve our understanding of the dynamics inherent in these generated evolving graphs.
\end{abstract}

\section*{Introduction}

Dynamic graphs refer to graphs subject to changes along time. 
Apart from the term 'dynamic graphs,' which can be found in \cite{harary.gupta.1997}, the terminology is varied. 
The most common terms mentioned in the scientific literature include 'evolving graphs' \cite{ferreira_note_2002}, 'dynamic networks' \cite{xuan.et.al.2003}, 'temporal networks' \cite{holme.saramaki.2012}, 'time-varying graphs' \cite{casteigts_time-varying_2011}, and 'temporal graphs' \cite{kostakos.2009}.
A Dynamic graph can be defined as a sequence of snapshot graphs ordered by a timestamp.
Many problems arising in a wide variety of systems have been formulated using dynamic graphs. 
Among them, as mentioned in \cite{boccaletti.et.al.2006}, the analysis and understanding of complex networks require the design of network growth models and graph evolution mechanisms.
The generation process always starts from an initial seed graph $G_0$ (the initial element of the sequence of snapshot graphs).
Then, at each step, a new graph is generated by applying rules to previously generated graphs.
A comprehensive description of this process is given in \cite{bridonneau_2023}. 
This new graph is then appended to the sequence, and the process continues until a specified condition is met or results in an infinite number of graphs.
The snapshot graph produced at step $t$ is both the current last element of the sequence produced by the generator and a resource element for the generator itself as illustrated on Figure \ref{fig:generator}.

\begin{center}
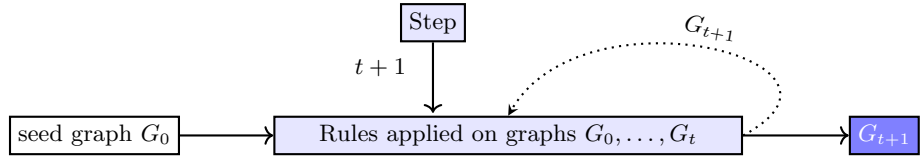
\begin{figure}[ht]
\centering
\begin{tikzpicture}[-,thick,xscale=1.0,yscale=1.0]
\tikzstyle{process}=[rectangle,draw=black,fill=blue!10,text=black]
\tikzstyle{input}=[rectangle,draw=black,fill=white,text=black]
\tikzstyle{output}=[rectangle,draw=black,fill=blue!50,text=white]

    \node[process,text width=6cm,align=center] (G) at (6,5) {Rules applied on graphs $G_0, \dots, G_t$};
    \node[input] (G0) at (0.5,5) {seed graph $G_0$};
    \node[process] (CK) at (5,6.5) {Step};
    \node (TIME) at (4.3,5.9) {$t+1$};
    \node[output] (TN) at (11,5) {$G_{t+1}$};
    
    \draw[->,thick] (G0) -- (G);
    \draw[->,thick] (CK) -- (5,5.3);
    \draw[->,thick] (G) -- (TN);
    \draw[->,dotted,>=stealth,thick] (G.east) .. controls (11,6) and (7,7) .. (G.north)
        node[sloped, above,midway] {$G_{t+1}$};
\end{tikzpicture}
\caption{Synthetic description of Dynamic Graphs Generators}
\label{fig:generator}
\end{figure}
\end{center}

Numerous challenges in graph theory have been revisited in the context of dynamic graphs. By introducing the temporal dimension, novel metrics have emerged, and classical properties have been redefined. These include time-respecting paths, reachability, temporal connectivity, and persistent patterns, among others. 
Nevertheless, it is noteworthy that, to the best of our knowledge, limited research has been dedicated to exploring the intricate relationship between the generative process and the inherent dynamics of the graph itself.
Some prior works presuppose a limited number of changes between two consecutive snapshot graphs. 
Others characterize the dynamics using terms like 'slow dynamics', 'not frequent changes', 'similar consecutive snapshot graphs' to cite a few.  
In both scenarios, there exists a clear need for a metric capable of quantifying the dynamism of the graph between two consecutive snapshot graphs. This metric should not only capture changes in the cardinality of vertex and edge sets but also alterations in their composition.

In this work, the DynamicScore metric, coping with both aspects, is presented\footnote{In \cite{bridonneau_2023}, this metric was referred to as 'nervousness,' a translation of a French term that could be misleading in English.}.
The metric is implemented for the set of vertices, V-DynamicScore, and for the set of edges, E-DynamicScore.
Our main contribution is a novel analysis of two state-of-the-art dynamic graphs generative processes based on this metric: the preferential attachment growing model by Barab{\'a}si and Albert \cite{barabasi.albert.1999} and the Edge-Markovian Graph model \cite{clementi.et.al.2010}.
In the next Section the metric is formally defined and some singular values corresponding to peculiar graph evolutions are presented and discussed.  
Section \ref{sec:PAandDS} is dedicated to the analysis of DynamicScore on graphs generated using the Preferential Attachment growing model. 
It is shown that the dynamics of the graph decreases as the number of steps increases.
Section \ref{sec:emg} starts with a description of the Edge-Markovian Graphs Generator (EMGG) and outlines some properties of the generated graphs. 
Then the analysis of the dynamics of Edge-Markovian graphs is conducted and some results about edge dynamics with respect to the parameters of the model are presented. 
We conclude this work by introducing two open questions about relationships between Markovian-based dynamic graph evolution and the DynamicScore.

\section{DynamicScore}

The DynamicScore, which is derived from the Jaccard distance, encompasses several properties that shed light on the nature of a dynamic graph. 
It effectively captures the degree of dynamics exhibited by the graph, whether it is applied to the vertices or the edges. 
Notably, DynamicScore emphasizes changes in composition, both at a local level over time between two consecutive steps, and at a global level spanning the entire graph.
It is formally defined as follow: 
\begin{definition}
\underline{V-DynamicScore}:\\
Given a dynamic graph $G$, such that at time $t$ $G_t = (V_t,E_t)$.
We call {\bf V-DynamicScore} at time $t$ and denoted by $\mathcal{D}^v_t$, the ratio:
\[
  \mathcal{D}^v_t = \frac{|V_{t+1} \triangle V_t|}{|V_{t+1} \cup V_t|}
\]
where $|A|$ denotes the number of edges present in set $A$. The $\triangle$ operator for all set $A$ and $B$, referred to as $A\triangle B$, is defined as $A \cup B - A \cap B$.
\end{definition}
Similarly, for a given dynamic graph the definition of its edges DynamicScore is defined as follow:
\begin{definition}
\underline{E-DynamicScore}:\\ Given a dynamic graph $G$, such that at time $t$ $G_t = (V_t,E_t)$. 
We call {\bf E-DynamicScore} at time $t$ and denoted by $\mathcal{D}^e_t$, the ratio:
\[
  \mathcal{D}^e_t = \frac{|E_{t+1} \triangle E_t|}{|E_{t+1} \cup E_t|}
\]
\end{definition}
The DynamicScore serves as a similarity metric, enabling comparisons between two consecutive snapshot graphs.
A score of 0 indicates that the two graphs are identical, while a score of 1 signifies that they do not share any common vertices.
In general, a value close to 0 suggests minimal changes in the graph between two consecutive steps, whereas a value close to 1 implies significant modifications have occurred.
It should be noted that graph order and DynamicScore measure two different quantities.
For instance, between two consecutive time steps, $t$ and $t+1$, the value of Vertex-DynamicScore can be equal to 1 while the order of the graph remains the same. 
This occurs when all the vertices have changed between $t$ and $t+1$. 
In the next two sections the analysis will mainly focus on the dynamics of Vertex and E-DynamicScore of the Barabas\'{i} model as defined in \cite{barabasi.albert.1999} and the EMGG model.

\section{Analysis of the Dynamics of the Preferential Attachment Growing Model}
\label{sec:PAandDS}

\subsection{Introduction to the Model}

In \cite{barabasi.albert.1999}, the generative process is clearly described.
For the first part of our analysis, we only focus on the evolution of the number of vertices and on the number of edges. 
Using our notations the generation of the graph starts with a seed graph $G_0 = (V_0,E_0)$ such that $|V_0| = n_0$ and $0 \leq |E_0| = m_0 \leq \frac{1}{2}n_0(n_0-1)$. Note that in the original research article, no information is given about the initial number of edges.
At every time step $t+1$ a new vertex is added and this new vertex is linked to $m (\leq n_0)$ vertices already in $V_t$.
Thus $|V_{t+1}| = |V_t|+1$ and $|E_{t+1}| = |E_t|+m$.

 
\subsection{DynamicScore}
From this it is possible to compute both Vertex and E-DynamicScore.
As the number of node inserted in the graph at each step is one, $\mathcal{D}^v_t=\frac{1}{n_0+t+1}$.
Moreover, the number of new connections being $m$ and no connection being removed leads to $\mathcal{D}^e_t=\frac{m}{m_0+tm}$.
Thus, both the Vertex and the E-DynamicScore are decreasing and tends toward 0 as $t$ tends to infinity.

\section{Generator of Edge-Markovian Graphs\label{sec:emg}}

This section presents the Edge-Markovian Graphs Generator (EMGG), its formal definition and some of its fundamental properties. 
In the first part, we present the model and its characteristics.
Moving on to the second part, we delve into the general results and explore the relationships between EMGG and the DynamicScore metric. 
These results unveil a significant connection between the graph's density and the value of DynamicScore, shedding light on their interplay and implications. 
By examining this relationship, we gain valuable insights into the dynamics of the graph and the quantitative assessment provided by DynamicScore.

\subsection{The Model}

The Edge-Markovian Graphs Generator (EMGG) is a stochastic process that produces an infinite sequence of static graphs. 
We denote $G_t$ the graph produced at step $t$.
$G_t = (V_t,E_t)$ where $V_t$ (resp. $E_t$) represents the set of vertices (resp. edges) at step $t$.

The EMGG is parameterized by two probabilities, denoted as $p$ and $q$, along with an initial condition or seed graph, denoted as $G_0$.
The set of vertices of the graph does not change during the evolution process, so, for all $t>0$, $V_t = V_0 = n$.
Given two vertices $u$ and $v$, if at step $t$ the edge $(u,v) \in E_t$, the edge is said {\it present} and {\it absent} otherwise. 
The EMGG operates as follows: at each step, all possible edges (present or absent) are examined\footnote{there are $n(n-1)/2$ such edges}. 
The generator determines for each edge if it will remain in the same state (present/absent) in the next snapshot graph or if it will change.
The decision is based on two probability parameters: $0 \leq p \leq 1$ and $0 \leq q \leq 1$.
The role of $p$ is to define the probability that an edge present at a given step remains present during the next step, while the role of $q$ is to define the probability that an edge absent at a given step remains absent during the next step.
This is summarized in the following diagram:
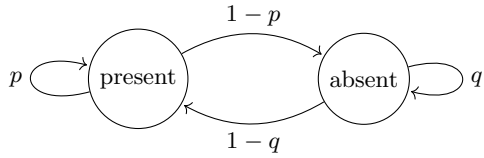
\begin{figure}[ht]
    \centering
    \begin{tikzpicture}[->,shorten >=1pt,node distance=3cm,on grid,auto]
       \node[state] (A) {present};
       \node[state] (B) [right=of A] {absent}; 
       \path[->]
        (A) edge [loop left] node {$p$} (A)
            edge [bend left] node {$1-p$} (B)
        (B) edge [loop right] node {$q$} (B)
            edge [bend left] node {$1-q$} (A);
    \end{tikzpicture}
    \caption{Description of the states.}
    \label{fig:enter-label}
\end{figure}

There are several special cases worth noting. 
When both $p$ and $q$ are set to 0, the generated graphs exhibit a blinking behavior, where edges alternate between present and absent at each step. 
On the other hand, when both $p$ and $q$ are set to 1, the generated graphs remain static throughout the sequence, with $G_t$ being equal to the initial graph $G_0$ for all time steps. 
In the case where $q = 1-p$, the generating process becomes "time-homogeneous", meaning that the generation of the new graph at each step does not depend on the previous step.

\begin{definition}
An EMGG is parameterized through 4 parameters $n \in \mathbb{N}^*$, $p$ and $q \in [0, 1]$ and an initial configuration $G_0$.
Instances produced by such a generator are such that:
\begin{itemize}
\item for all step $t$, $|V_t|=n$;
\item for pair of vertices $e = (u, v) \in V^2_t$:
\begin{itemize}
    \item if $e \in E_t$, then $e \in E_{t+1}$ (remain present) with probability $p$ and becomes absent with probability $1-p$;
    \item if $e \notin E_t$, then $e \in E_{t+1}$ (becomes present) with probability $1-q$ and remain absent of $E_{t+1}$ with probability $q$.
\end{itemize}
\end{itemize}
\end{definition}
The maximum number of edges that may be contained at a given step $t$ is $\binom{n}{2}$.
The set of edges is evolving through time and computing E-DynamicScore gives an information about its dynamics. 
In the following sub sections we establish a solid foundation for understanding its dynamics and its relationship with probabilities $p$ and $q$.\\
\textbf{Note:} in the following, the number of edges in a generated graph at step $t$ will be referred to as $m_t$ and the graph density will be referred to as $\hat{m_t}$.

\subsection{Known Properties of EMGG}

In order to ease the understanding of the dynamics of EMGG instances, some results about EMGG are presented.

First note that the state of each edge is independent of the state of the other edges of the graph, thus, studying the probability of presence/absence of each edge independently from the others is correct. 
As presented in \cite{clementi.et.al.2010} the transition matrix $P$ for a single edge satisfies:
\begin{equation}
\label{eq:matrixPTransition}
P = \begin{pmatrix}
    p & 1-p \\
    1-q & q
\end{pmatrix}
\end{equation}
The analysis of Markovian processes and more especially the study of two-states markovian processes has shown that for each single edge, the distribution of presence, in the context of EMGG, converges toward a stationary distribution $\pi$ as long as $|p+q-1| \neq 1$. 
The situation $|p+q-1| = 1$ is discussed after the proof of the theorem. 
As a stationary distribution of a Markov chain, $\pi$ satisfies $\pi=\pi P$.
The value of vector $\pi$ is stated in the following theorem:
\begin{theorem} \underline{Stationary distribution}:\\
For $p, q$ probabilities such that $|p + q - 1| \neq 1$, the stationary distribution $\pi$ is $\left(\frac{1-p}{2-p-q} ~~ \frac{1-q}{2-p-q}\right)$.
\end{theorem}
\noindent \textbf{Proof}:
It is sufficient to notice that $\pi = \left(\frac{1-p}{2-p-q} \frac{1-q}{2-p-q}\right)$ is a distribution and that $\pi = \pi P$.
\qed

Thus, the presence of an edge has a Bernoulli distribution of parameter 
$\pi^* = \frac{1-p}{2-p-q}$ as a stationary distribution.
As every edge is independent one from the other, the number of edges has a binomial distribution of parameter $\binom{n}{2}$ and $\pi^*$.
The situation for which $|p+q-1| = 1$ as two subcases, either $p=q=1$ or $p=q=0$.
On the one hand if $p=q=1$, then graphs produced by EMGG remains unchanged forever.
This means $G_t = G_0$ for all $t$.
On the other hand if $p=q=0$, then produced graphs are 2-periodic and more precisely, $E_{t+1}$ is the complementary of $E_t$ for all $t$.
Thus, in both cases the density of a produced graph does not converge to a stationary distribution.

\subsection{EMGG and E-DynamicScore}

This subsection presents several key results concerning the Edge-Markovian Graphs Generator (EMGG) and the E-DynamicScore of the graphs it generates.
Results stated here are specific cases of the analysis made in the previous section.
Every result mentioned in this section will be connected to ones stated above.
Firstly, we provide the computation of the density of these graphs, a fundamental quantity in the context of EMGG.
The expectation of this quantity is then stated, offering insights into its average behavior.
Moving forward, we examine the average DynamicScore across all possible density values. 
By analyzing this metric, we gain a comprehensive understanding of the dynamics of the EMGG and its relationship with the density parameter. 
Specifically, we explore the DynamicScore at the fixed point density, uncovering the crucial role played by the probabilities $p$ and $q$, and elucidating the characteristics that can be derived from this special value. 
Notably, we establish a meaningful connection between this particular value of the DynamicScore and the values obtained through experimental observations.
Through these results, we deepen our understanding of the EMGG and its association with E-DynamicScore, providing valuable insights into the dynamics and quantitative assessment of this stochastic graph generation process.\\

\noindent \textbf{Density Evolution of Edge-Markovian Graphs:}

In order to better understand the relationships between EMGG dynamics and the E-DynamicScore metric, we first show 
that the number of edges is on average close to a quantity depending only on $p$ and $q$.
To that end, we prove the following lemma on the evolution of the density:
\begin{lemma}
\underline{Evolution of the Density}\\
Let consider $EMGG$ parameterized by $(n, p, q)$.
Let $(G_0, \dots, G_t)$ be a sequence of graphs produced by EMGG.
Then, the expected normalized density for the graph $G_{t+1}$ satisfies the following equation:
\begin{equation}
    \hat{m_{t+1}} \simeq f_{p, q}(\hat{m_t}) = \hat{m_t}p + (1-q)(1-\hat{m_t}) = (p + q - 1)\hat{m_t}
\end{equation}
\end{lemma}
\textbf{Proof:} as the process is a Markov chain, $\hat{m_{t+1}}$ depends only on $\hat{m_t}$.
Second, it is worth mentioning that every edge is independent from the others.
The expected number of edges that remain present is $p\hat{m_t}$ while the expected number of edges changing their state from absent to present is $(1-q)(1-\hat{m_t})$.
The expected number of edges present at step $t+1$ is thus the sum of these two quantities. \qed

This lemma provides a valuable interpretation of the density expectation, which allows us to further investigate the existence of a fixed density.
By analyzing the expectation, we can identify a specific value that represents a fixed point within the computation process.
In the context of the function $f_{p,q}$, a fixed point refers to a value $m^*$ for which $f(m^*)=m^*$ holds true.
The computation of this fixed point value is carried out according to the procedure outlined in the subsequent lemma.
\begin{lemma}
\label{lem:ExpNE}
\underline{Expected Number of Edges:}\\
Let $G$ be a graph produced by $EMG(n, p, q)$
Let $\hat{m_t}$ be the density of graph at step $t$.
Then, as long as $|p + q - 1| < 1$ an expectation value for $\hat{m_t}$, referred to as $m^*$, satisfies $f_{p,q}(m^*)=m^*$:
\begin{equation}
    m^* = \frac{1-q}{2-p-q}
\end{equation}
\end{lemma}
\textbf{Proof:} this result comes from finding a fixed point to the function $f_{p,q}$ \qed

This fixed point value matches with the probability of presence of an active edge in the stationary regime.
It is not surprising as it gives, in both case, the average and expected value of the graph density.
These findings enable us to gain deeper insights into the dynamics of the system and the properties associated with the EMGG, paving the way for a more comprehensive understanding of its behavior.


\subsection{Relationship with the DynamicScore}

This section explores the relationship between the Edge-Markovian Graphs Generator (EMGG) and the DynamicScore, focusing on the computation of an expectation value regardless of the graph's density.
The following theorem provides a precise value of this expectation, elucidating the crucial role played by the parameters $p$ and $q$ in this context:
\begin{theorem}
\label{theo:AvgGeneralDynScore}
\underline{Average General DynamicScore}\\
Let $G$ be a graph produced by $EMG(n, p, q)$
Let $\hat{m_t}$ be the density of graph at step $t$.
Then, in average:
\begin{equation}
\label{eq:nerv_EMG}
\mathcal{D}^E_t = 1-\frac{p\hat{m_t}}{1+q(\hat{m_t} - 1)}
\end{equation}
\end{theorem}
\textbf{Proof:} The proof consists in finding the average number of edges in $E_t \triangle E_{t+1}$ and in $E_t \cup E_{t+1}$.
For the first one, it consists in computing, on average, the number of edges which state is changing. 
Assuming the density of edges at $t$ is $\Hat{m_t}$, then the density of edges that change from present to absent is on average $(1-p)\Hat{m_t}$ and the density of newly present edges is on average $(1-q)(1-\Hat{m_t})$. Therefore, the size of $E_t \triangle E_{t+1}$ is on average:
\[
|E_t \triangle E_{t+1}| = (1-p)\Hat{m_t} + (1-q)(1-\Hat{m_t})
\]
For computing the union size, it is sufficient to notice that it contains all the present edges at step $t$ plus appearing edges $(1-q)(1-\Hat{m_t})$.
Thus, the size of the union is in average:
\[
|E_t \cup E_{t+1}| = \Hat{m_t} + (1-q)(1-\Hat{m_t})
\]
It is therefore possible to estimate the average DynamicScore:
\begin{equation*}
    \mathcal{D}^E_t = \frac{(1-p)\Hat{m_t} + (1-q)(1-\Hat{m_t})}{\Hat{m_t} + (1-q)(1-\Hat{m_t})} = 1-\frac{p\Hat{m_t}}{1 + q(\Hat{m_t} - 1)}
\end{equation*}\qed

This result must be evaluated for densities close to $m^*$.
Indeed, the distribution of the edges follows a binomial law of parameters $\binom{n}{2}$ and $\pi^*$.
Therefore most values of $|E_t|$ taken by generated graphs are close to the expected value of the binomial law: $\binom{n}{2}\pi^*$.
This implies density of these graphs are close to $\pi^*=m^*$.
Combining this theorem with the fixed point density of generated graphs provides DynamicScore at the fixed point density:
\begin{theorem}
\underline{E-DynamicScore in Average around $m^*$:}\\
For all $p,q$ such that $|p + q - 1| < 1$
\[
\mathcal{D}^E_t(m^*) = 2\frac{1-p}{2-p}
\]
Moreover, $\mathcal{D}^E_t(m^*)$ may take all the values from 0 to 1.
\end{theorem}
\textbf{Proof:} it results from the combination of both theorem \ref{theo:AvgGeneralDynScore} and lemma \ref{lem:ExpNE}. \qed

Notably, the average DynamicScore is independent of the value of $q$, and it exhibits a decreasing trend as $p$ increases.
The range of possible values for the DynamicScore ranges from 0 to 1, indicating its ability to capture the extent of changes in the graph.
To illustrate these findings, several figures are presented.
These figures have been obtained through simulations, considering various values of $p$ and $q$, both ranging from 0 to 1, while excluding the endpoints.
These visual representations offer a good understanding of the relationship between EMGG instances, their DynamicScore on average, and the parameters $p$ and $q$.
By examining these figures, we obtain experimental confirmation and deeper understanding of the behavior and characteristics of the EMGG, corroborating the insights provided by the above-stated theorem, particularly in relation to the DynamicScore.
The impact of the parameter $q$ on the average DynamicScore is found to be negligible, whereas parameter $p$ appears to be more influential in determining its value.
Notably, it is observed that the DynamicScore can encompass the entire range from 0 to 1 as $p$ varies from 1 to 0.

\begin{figure}[h]
    \centering
    \begin{subfigure}{0.4\textwidth}
        \centering
        \begin{tikzpicture}
            \begin{axis}[
                xlabel={$q$},
                ylabel={${\mathcal{D}^E}_t(m^*)$},
                width=\textwidth,
                height=6.5cm,
                grid=major,
                only marks,
                legend style={
                    font=\scriptsize,
                    anchor=north west,
                },
                cycle list name=exotic
            ]
            
            \foreach \p in {0.05, 0.15, 0.25, 0.35, 0.45, 0.55, 0.65, 0.75, 0.85, 0.95} {
                \addplot +[mark=*, mark size=2pt, y filter/.expression={\thisrow{p} == \p ? \thisrow{nervousnessAverage} : nan}] table [x=q, y=nervousnessAverage, col sep=comma] {data_150_nervousness_average_no_id.csv};
                \addlegendentryexpanded{$p =${\p}}
            }
            
            \end{axis}
        \end{tikzpicture}
        \caption{E-DynamicScore Average vs. $q$.}
    \end{subfigure}
    \hspace{2cm}
    \hfill
    \begin{subfigure}{0.4\textwidth}
        \centering
        \begin{tikzpicture}
            \begin{axis}[
                xlabel={$p$},
                ylabel={${\mathcal{D}^E}_t(m^*)$},
                width=\textwidth,
                height=6.5cm,
                grid=major,
                only marks,
                legend style={
                    font=\scriptsize,
                    anchor=north west,
                },
                cycle list name=exotic
            ]
            
            \foreach \q in {0.05, 0.15, 0.25, 0.35, 0.45, 0.55, 0.65, 0.75, 0.85, 0.95} {
                \addplot +[mark=*, mark size=2pt, y filter/.expression={(\thisrow{q} == \q) ? \thisrow{nervousnessAverage} : nan}] table [x=p, y=nervousnessAverage, col sep=comma] {data_150_nervousness_average_no_id.csv};
                \addlegendentryexpanded{$q = \q$}
            }
            
            \end{axis}
        \end{tikzpicture}
        \caption{E-DynamicScore Average vs. $p$}
    \end{subfigure}
    \caption{Average dynamic score as a function of the parameters $p$ and $q$.
    On the left, the parameter $p$ is set and the parameter $q$ ranges from 0.05 to 0.95.
    One may notice that for a fixed value of parameter $p$, the average dynamicScore does not depend on $q$.
    On the right, the parameter $q$ is set and the parameter $p$ ranges from 0.05 to 0.95.
    As observed with the picture on the left, the average dynamicScore does not depend on $q$ so all the marks are mingled.}
\end{figure}
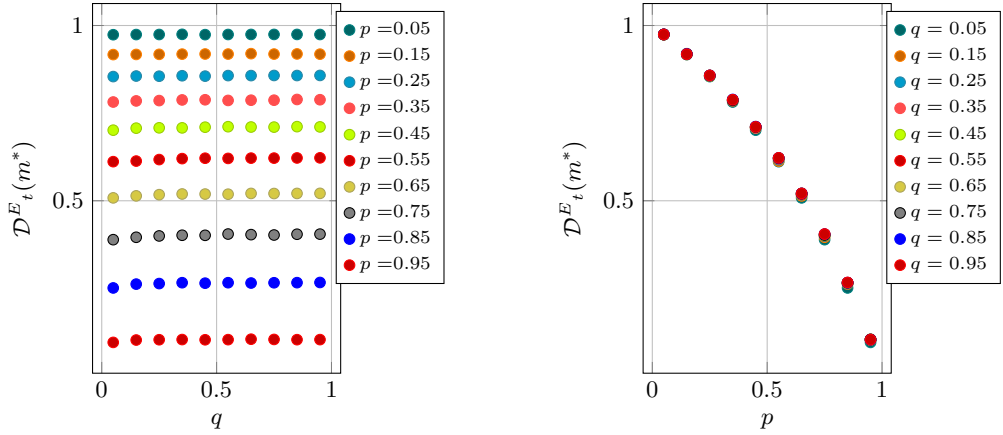
%
%

\section{Conclusion and Open Problems}

In this work, a new metric called DynamicScore has been presented. 
This metric, proposed for both edges and vertices, quantifies the evolution of the dynamics of dynamic graphs. 
It has been demonstrated that the Preferential Attachment growing model generates graphs with dynamics that tends toward zero.
This implies that after numerous iterations, the dynamic graph undergoes minimal changes, resulting in a stability of the properties within the generated graphs.
However, the dynamics of real complex networks is not solely reliant on the creation of vertices and edges but also on the removal of vertices and edges.
This leads us to the following open question: (i) {\it given a specific dynamic graphs generator, is there a relationship between DynamicScore values and the preservation of properties in dynamic graphs?} 

The second studied generator was the Edge-Markovian Graphs Generator. 
The mechanics of this generator is based on two probabilistic parameters, $p$ and $q$, driving the states of edges that can be present or absent.  
After an in-depth analysis of the average density of the generated graphs, using DynamicScore, it has been shown that the dynamics of such graphs is only driven by probability parameter $p$. 
The analysis relies on the Markovian nature of the generator, which prompts two additional open questions: (ii) {\it if the evolution/generative process exhibits Markovian characteristics in the evolution of edges, does the value of E-DynamicScore remain nearly constant?} and (iii) {\it conversely, if the value of DynamicScore remains constant, does this indicate that the evolution process is Markovian?}

\bibliographystyle{plainurl}
\bibliography{biblio}   

\end{document}